\PassOptionsToPackage{unicode}{hyperref}
\PassOptionsToPackage{hyphens}{url}
\PassOptionsToPackage{dvipsnames,svgnames,x11names}{xcolor}
\documentclass[
  letterpaper,
  DIV=11,
  numbers=noendperiod]{scrartcl}

\usepackage{amsmath,amssymb}
\usepackage{lmodern}
\usepackage{iftex}
\ifPDFTeX
  \usepackage[T1]{fontenc}
  \usepackage[utf8]{inputenc}
  \usepackage{textcomp} 
\else 
  \usepackage{unicode-math}
  \defaultfontfeatures{Scale=MatchLowercase}
  \defaultfontfeatures[\rmfamily]{Ligatures=TeX,Scale=1}
\fi
\IfFileExists{upquote.sty}{\usepackage{upquote}}{}
\IfFileExists{microtype.sty}{
  \usepackage[]{microtype}
  \UseMicrotypeSet[protrusion]{basicmath} 
}{}
\makeatletter
\@ifundefined{KOMAClassName}{
  \IfFileExists{parskip.sty}{%
    \usepackage{parskip}
  }{
    \setlength{\parindent}{0pt}
    \setlength{\parskip}{6pt plus 2pt minus 1pt}}
}{
  \KOMAoptions{parskip=half}}
\makeatother
\usepackage{xcolor}
\ifx\paragraph\undefined\else
  \let\oldparagraph\paragraph
  \renewcommand{\paragraph}[1]{\oldparagraph{#1}\mbox{}}
\fi
\ifx\subparagraph\undefined\else
  \let\oldsubparagraph\subparagraph
  \renewcommand{\subparagraph}[1]{\oldsubparagraph{#1}\mbox{}}
\fi

\usepackage{longtable,booktabs,array}
\usepackage{calc} 
\usepackage{etoolbox}
\makeatletter
\patchcmd\longtable{\par}{\if@noskipsec\mbox{}\fi\par}{}{}
\makeatother
\IfFileExists{footnotehyper.sty}{\usepackage{footnotehyper}}{\usepackage{footnote}}
\makesavenoteenv{longtable}
\usepackage{graphicx}
\makeatletter
\def\maxwidth{\ifdim\Gin@nat@width>\linewidth\linewidth\else\Gin@nat@width\fi}
\def\maxheight{\ifdim\Gin@nat@height>\textheight\textheight\else\Gin@nat@height\fi}
\makeatother
\setkeys{Gin}{width=\maxwidth,height=\maxheight,keepaspectratio}
\makeatletter
\def\fps@figure{htbp}
\makeatother
\newlength{\cslhangindent}
\newlength{\csllabelwidth}
\newlength{\cslentryspacingunit} 
\newenvironment{CSLReferences}[2] 
 {
  \setlength{\parindent}{0pt}
  \ifodd #1
  \let\oldpar\par
  \def\par{\hangindent=\cslhangindent\oldpar}
  \fi
  \setlength{\parskip}{#2\cslentryspacingunit}
 }%
 {}
\usepackage{calc}

\KOMAoption{captions}{tableheading}
\makeatletter
\@ifpackageloaded{caption}{}{\usepackage{caption}}
\AtBeginDocument{%
\ifdefined\contentsname
  \renewcommand*\contentsname{Table of contents}
\else
  \newcommand\contentsname{Table of contents}
\fi
\ifdefined\listfigurename
  \renewcommand*\listfigurename{List of Figures}
\else
  \newcommand\listfigurename{List of Figures}
\fi
\ifdefined\listtablename
  \renewcommand*\listtablename{List of Tables}
\else
  \newcommand\listtablename{List of Tables}
\fi
\ifdefined\figurename
  \renewcommand*\figurename{Figure}
\else
  \newcommand\figurename{Figure}
\fi
\ifdefined\tablename
  \renewcommand*\tablename{Table}
\else
  \newcommand\tablename{Table}
\fi
}
\@ifpackageloaded{float}{}{\usepackage{float}}
\floatstyle{ruled}
\@ifundefined{c@chapter}{\newfloat{codelisting}{h}{lop}}{\newfloat{codelisting}{h}{lop}[chapter]}
\floatname{codelisting}{Listing}

\makeatother
\makeatletter
\@ifpackageloaded{caption}{}{\usepackage{caption}}
\@ifpackageloaded{subcaption}{}{\usepackage{subcaption}}
\makeatother
\makeatletter
\@ifpackageloaded{tcolorbox}{}{\usepackage[many]{tcolorbox}}
\makeatother
\makeatletter
\@ifundefined{shadecolor}{\definecolor{shadecolor}{rgb}{.97, .97, .97}}
\makeatother
\ifLuaTeX
  \usepackage{selnolig}  
\fi
\IfFileExists{bookmark.sty}{\usepackage{bookmark}}{\usepackage{hyperref}}
\IfFileExists{xurl.sty}{\usepackage{xurl}}{} 
\hypersetup{
  pdftitle={Generative Geostatistical Modeling from Incomplete Well and Imaged Seismic Observations with Diffusion Models},
  pdfauthor={Huseyin Tuna Erdinc1, Rafael Orozco1, and Felix J. Herrmann1 1 Georgia Institute of Technology},
  colorlinks=true,
  linkcolor={blue},
  filecolor={Maroon},
  citecolor={Blue},
  urlcolor={Blue},
  pdfcreator={LaTeX via pandoc}}

\title{Generative Geostatistical Modeling from Incomplete Well and
Imaged Seismic Observations with Diffusion Models}
\author{Huseyin Tuna Erdinc\textsuperscript{1}, Rafael
Orozco\textsuperscript{1}, and Felix J. Herrmann\textsuperscript{1}\\
\textsuperscript{1} Georgia Institute of Technology}
\date{}
\begin{document}
\maketitle


\hypertarget{abstract}{%
\subsection*{Abstract}\label{abstract}}
\addcontentsline{toc}{subsection}{Abstract}

In this study, we introduce a novel approach to synthesizing subsurface
velocity models using diffusion generative models. Conventional methods
rely on extensive, high-quality datasets, which are often inaccessible
in subsurface applications. Our method leverages incomplete well and
seismic observations to produce high-fidelity velocity samples without
requiring fully sampled training datasets. The results demonstrate that
our generative model accurately captures long-range structures, aligns
with ground-truth velocity models, achieves high Structural Similarity
Index (SSIM) scores, and provides meaningful uncertainty estimations.
This approach facilitates realistic subsurface velocity synthesis,
offering valuable inputs for full-waveform inversion and enhancing
seismic-based subsurface modeling.

\hypertarget{introduction}{%
\subsection{Introduction}\label{introduction}}

Diffusion generative models are powerful frameworks for learning
high-dimensional distributions and synthesizing high-fidelity images.
However, their efficacy in training predominantly hinges on the
availability of complete, high-quality training datasets, a condition
that often proves unattainable, particularly in the domain of subsurface
velocity-model generation. In this work, we propose to synthesize proxy
subsurface velocities from incomplete well and imaged seismic
observations by introducing additional corruptions to the observations
during the training phase. In this context, proxy velocity models refer
to random realizations of subsurface velocities that are close in
distribution to the actual subsurface velocities. These proxy models can
be used as priors to train neural networks with simulation-based
inference. Our approach facilitates the generation of these proxy
velocity samples by utilizing available datasets composed merely of
seismic images and 5 (for now) wells per seismic image. After training,
our generative model permits the generation of velocity
samples derived from unseen RTMs without the need of having access to
wells.

\begin{figure}[t]

{\centering 

\includegraphics{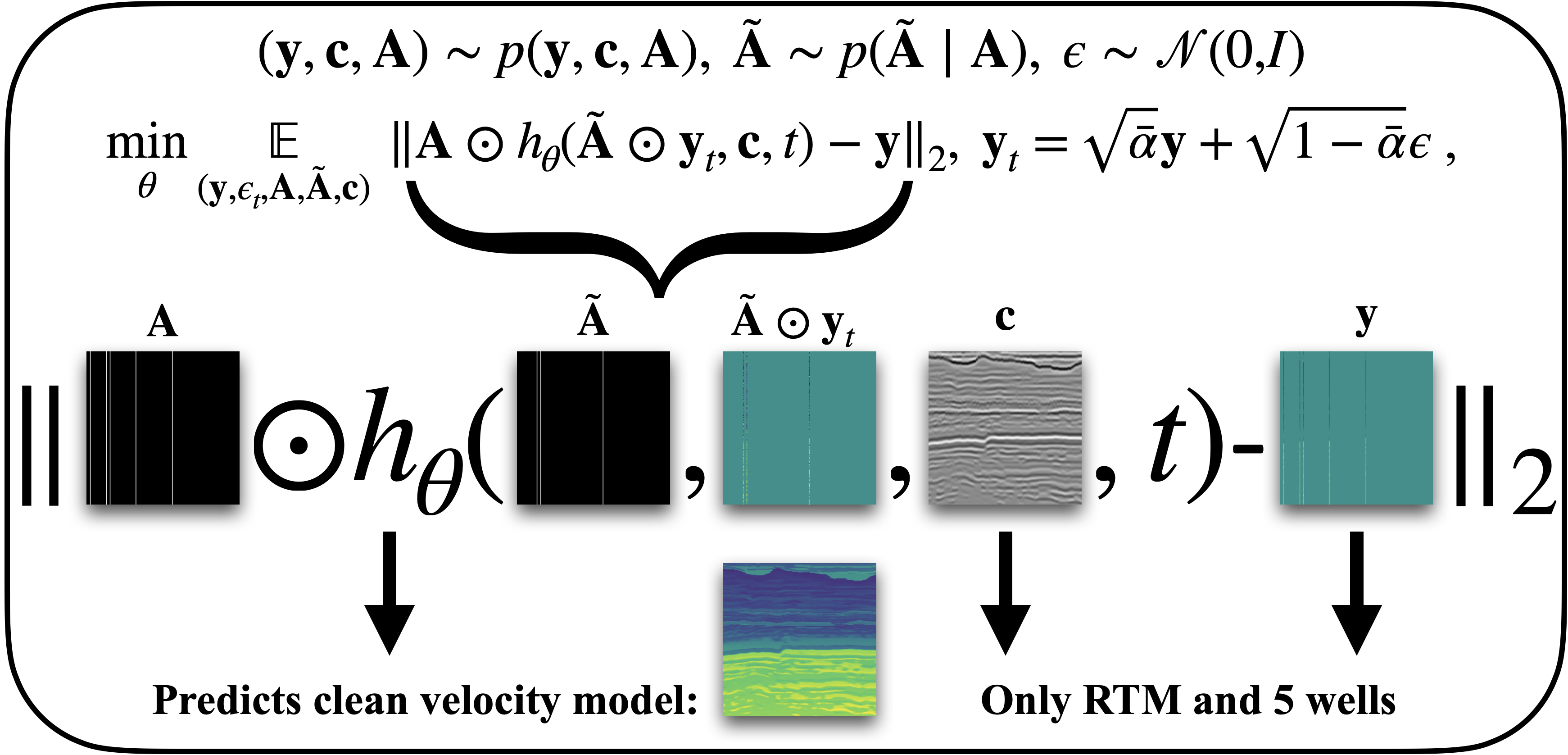}

}

\caption{\label{fig-sup}Training-loss formulation: field observations
(RTMs and wells) along with column-wise masks (\(\mathbf{A}\)) are
sampled initially. Conditioned on these masks, further sub-sampling
masks (\(\mathbf{\widetilde{A}}\)) are generated. The neural denoising
model \(h_\theta\), informed by the sub-sampling masks, noisy
sub-sampled well information and RTMs, is trained to reconstruct the
complete, noise-free velocity model.}

\end{figure}

\hypertarget{methodology}{%
\subsection{Methodology}\label{methodology}}

Seismic velocity-model synthesis from incomplete measurements with
Generative Geostatistical Modeling (GGM) is an ill-posed problem. In
addition, acquiring comprehensive realistic velocity-model training
datasets from seismic shot data through the process of full-waveform
inversion proves to be too costly. Generative models, particularly
diffusion models, offer a solution by training conditional neural
networks to synthesize plausible samples, that we refer to as proxy
models. After training, these proxy models are synthesized through an
iterative denoising process, facilitating high-fidelity,
high-dimensional sample generation. Our work leverages generative
diffusion models by producing proxy velocity models from limited
well-log and imaged seismic information. It negates the need of having
access to fully-sampled velocity models during training. Well data
provides the only incomplete access to the ground-truth velocities.

Unfortunately, this lack of fully-sampled velocity models precludes the
use of learning based methods that rely on having
access to fully-sampled training examples (Wang, Huang, and Alkhalifah
2024). Inspired by recent work by Giannis Daras and Klivans (2023),
Orozco, Louboutin, and Herrmann (2023) and Peters (2020), our study
adopts an alternative approach that only requires access to sub-sampled
velocity models (i.e.~well-log data, denoted by the sub-sampling
operator \(\mathbf{A}\)). The sub-sampling corresponds to the action of
binary masks with unity columns at locations where well-log data is
available. The reconstruction of the fully-sampled proxy velocity models
is conditioned on seismic images (denoted by \(\mathbf{c}\)). The
proposed training process is illustrated in Figure~\ref{fig-sup} and
repeatedly involves: selection of a 2D seismic image, \(\mathbf{c}\),
paired with 5 well logs \(\mathbf{y}\) from which the sub-sampling
masks, \(\mathbf{A}\) are generated. Conditioned on this sub-sampling
mask, a more sub-sampled mask, \(\mathbf{\widetilde{A}}\) (cf.~the masks
\(\mathbf{A}\) and \(\mathbf{\widetilde{A}}\) in Figure~\ref{fig-sup})
is generated by zeroing out more columns. Given the action of the more
sub-sampled mask on noised and scaled complete well-log data, the neural
network is trained to denoise and fit the complete well-log data
conditioned on the seismic image. Our network is trained by repeating
this process over 20,000 denoising iterations on 2 A100 GPUs involving a
training dataset of 3000 training samples.

\hypertarget{results}{%
\subsection{Results}\label{results}}

\begin{figure}[t]

{\centering 
\includegraphics{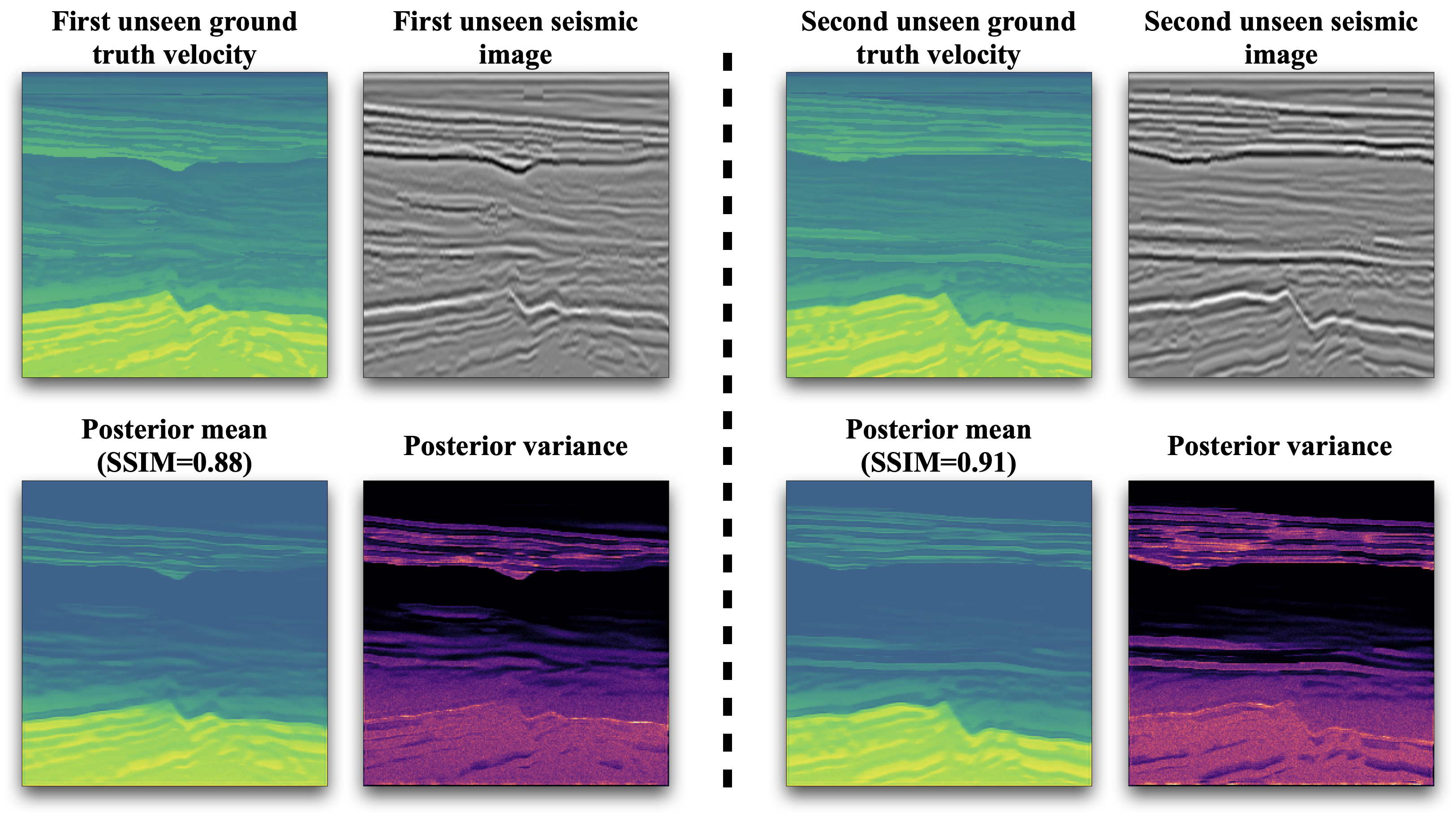}
}

\caption{\label{fig-unsup}The denoising generative model is tested on
two previously unseen seismic images. For each case, we fed the
corresponding seismic image into the network and generated velocity samples (200
samples for each example). Subsequently, we computed both the posterior
mean and variance. To assess the accuracy of our model, we included the
SSIM between the ground-truth to the posterior means}

\end{figure}

Figure~\ref{fig-unsup} presents results of our generative model for two
unseen examples of imaged reflectivities. Given these unseen seismic
images, our trained neural model synthesizes subsurface velocity
models without relying on well data. To validate the quality of our
generative samples, comparisons are made between the posterior means
(the average of 200 generative samples) and their respective unseen
ground-truth test velocity models. These velocity models remained
entirely unseen during the training and generative phase. We observe
that the posterior means can capture long range structures and produce
visually meaningful velocity samples, evidenced by the high Structural
Similarity Index (SSIM) scores of 0.88 and 0.91. Furthermore, the
conditional posterior variances align closely with particular structures
within the velocity models, which underscores the generative model's
capacity to accurately reflect the uncertainty due to two factors: the
variability in the subsurface prior information captured from wells seen
during training and the uncertainty due to the imaging process. From
these results, we conclude that our method conditioned on imaged seismic
data is capable of producing high-fidelity samples from the underlying
distribution for subsurface velocity models. These samples will serve as
input to our full-Waveform variational Inference via Subsurface
Extensions (WISE) framework (Yin et al. 2023). Looking ahead, future
directions include expanding our dataset across varied geological
settings, inclusion of prompts to guide the sample generation, all
geared towards advancing on the road to a seismic-based foundation model
for subsurface velocities.

\hypertarget{conclusion-and-significance}{%
\subsection{Conclusion and
Significance}\label{conclusion-and-significance}}

We utilized generative diffusion models to synthesize realistic
subsurface velocity model samples. Unlike prior approaches dependent on
fully sampled velocity model datasets, our method achieves training with
merely \(\approx 2\)\% of velocity information derived from well data
and corresponding seismic observations. After training is
completed, our model's sampling mechanism operates without well data,
conditioned solely on seismic images to generate velocity samples. The velocity
samples produced by our technique are highly beneficial for subsequent
tasks such as WISE, which rely on having access to high-fidelity samples
of the distribution of subsurface velocity models. Additional material
is available at https://slimgroup.github.io/IMAGE2024/.

\hypertarget{acknowledgement}{%
\subsection{Acknowledgement}\label{acknowledgement}}

This research was carried out with the support of Georgia Research
Alliance and partners of the ML4Seismic Center.

\hypertarget{references}{%
\subsection*{References}\label{references}}
\addcontentsline{toc}{subsection}{References}

\hypertarget{refs}{}
\begin{CSLReferences}{1}{0}
\leavevmode\vadjust pre{\hypertarget{ref-ambient}{}}%
Giannis Daras, Yuval Dagan, Kulin Shah, and Adam Klivans. 2023.
{``Ambient Diffusion: Learning Clean Distributions from Corrupted
Data.''} \emph{Advances in Neural Information Processing Systems
(NeurIPS)}.

\leavevmode\vadjust pre{\hypertarget{ref-rafaelkriging}{}}%
Orozco, Rafael, Mathias Louboutin, and Felix J. Herrmann. 2023.
{``Generative Seismic Kriging with Normalizing Flows.''} \emph{Third
International Meeting for Applied Geoscience and Energy (IMAGE)}.
\url{https://slimgroup.github.io/IMAGE2023/BayesianKrig/abstract.html}.

\leavevmode\vadjust pre{\hypertarget{ref-basdeep}{}}%
Peters, Bas. 2020. {``Shortcutting Inversion-Based Near-Surface
Characterization Workflows Using Deep Learning.''} \emph{SEG Technical
Program Expanded Abstracts 2020}.
\url{https://doi.org/10.1190/segam2020-3428431.1}.

\leavevmode\vadjust pre{\hypertarget{ref-wang2024controllable}{}}%
Wang, Fu, Xinquan Huang, and Tariq Alkhalifah. 2024. {``Controllable
Seismic Velocity Synthesis Using Generative Diffusion Models.''}
\url{https://doi.org/10.48550/arXiv.2402.06277}.

\leavevmode\vadjust pre{\hypertarget{ref-yin2023wise}{}}%
Yin, Ziyi, Rafael Orozco, Mathias Louboutin, and Felix J Herrmann. 2023.
{``WISE: Full-Waveform Variational Inference via Subsurface
Extensions.''} \emph{arXiv Preprint arXiv:2401.06230}.

\end{CSLReferences}

\end{document}